\documentclass[
reprint,
superscriptaddress,
aps,
prl
]{revtex4-2}

\usepackage{graphicx}
\usepackage{dcolumn}
\usepackage{bm}
\usepackage[utf8]{inputenc}
\usepackage{amsmath, amsthm, amssymb, amsfonts}
\usepackage[scr]{rsfso}
\usepackage{accents}
\usepackage[usenames,dvipsnames]{xcolor}
\usepackage{tikz}
\usepackage{array}
\usepackage{dsfont}
\usepackage{mathtools}
\usepackage{soul,xcolor}
\usepackage{enumerate}
\usepackage[colorlinks=true,linkcolor=Blue,citecolor=Blue,urlcolor=Blue]{hyperref}
\usepackage[caption=false]{subfig}
\usepackage[capitalise]{cleveref}

\newcommand{\YZ}{\color{black}}

\setstcolor{blue}

\begin{document}

\title{Unified treatment of synchronization patterns in generalized networks with higher-order, multilayer, and temporal interactions}

\author{Yuanzhao Zhang}
\email[]{yuanzhao@u.northwestern.edu}
\affiliation{Center for Applied Mathematics, Cornell University, Ithaca, New York 14853, USA}
\affiliation{Department of Physics and Astronomy, Northwestern University, Evanston, Illinois 60208, USA}
\author{Vito Latora}
\affiliation{School of Mathematical Sciences, Queen Mary University of London, London E1 4NS, United Kingdom}
\affiliation{Dipartimento di Fisica ed Astronomia, Università di Catania and INFN, I-95123 Catania, Italy}
\affiliation{The Alan Turing Institute, The British Library, London NW1 2DB, United Kingdom}
\affiliation{Complexity Science Hub Vienna (CSHV), Vienna, Austria}
\author{Adilson E. Motter}
\email[]{motter@northwestern.edu}
\affiliation{Department of Physics and Astronomy, Northwestern University, Evanston, Illinois 60208, USA}
\affiliation{Northwestern Institute on Complex Systems, Northwestern University, Evanston, Illinois 60208, USA}

\begin{abstract}
  When describing complex interconnected systems, one often has to go beyond the standard network description to account for generalized interactions. Here, we establish a unified framework to simplify the stability analysis of cluster synchronization patterns for a wide range of generalized networks, including hypergraphs, multilayer networks, and temporal networks. The framework is based on finding a simultaneous block diagonalization of the matrices encoding the synchronization pattern and the network topology. As an application, we use simultaneous block diagonalization to unveil an intriguing type of chimera states that appear only in the presence of higher-order interactions. The unified framework established here can be extended to other dynamical processes and can facilitate the discovery of emergent phenomena in complex systems with generalized interactions.

\vspace{3mm}
\noindent DOI: \href{https://doi.org/10.1038/s42005-021-00695-0}{10.1038/s42005-021-00695-0} 
\end{abstract}

\maketitle


Over the past two decades, networks have emerged as a versatile description of interconnected complex systems \cite{strogatz2001exploring,newman2003structure}, generating crucial insights into myriad social \cite{clauset2004finding}, biological \cite{bassett2011dynamic}, and physical \cite{motter2013spontaneous} systems.
However, it has also become increasingly clear that the original formulation of a static network representing a single type of pairwise interaction has its limitations.
For instance, neuronal networks change over time due to plasticity and comprise both chemical and electrical interaction pathways \cite{sporns2010networks}.
For this reason, the original formulation has been generalized in different directions, 
including {\it hypergraphs} that account for nonpairwise interactions involving three or more nodes simultaneously \cite{berge1973graphs,battiston2020networks}, {\it multilayer networks} that accommodate multiple types of interactions \cite{kivela2014multilayer,boccaletti2014structure}, and {\it temporal networks} whose connections change over time \cite{holme2012temporal}.
Naturally, with the increased descriptive power comes increased analytical complexity, especially for dynamical processes on these generalized networks.

One important class of dynamical processes on networks is cluster synchronization.
Many real-world networks show intricate cluster synchronization patterns, where one or more internally coherent but mutually independent clusters coexist \cite{stewart2003symmetry,belykh2008cluster,dahms2012cluster,nicosia2013remote,williams2013experimental,rosin2013control,fu2013topological,brady2021forget}. 
Maintaining the desired dynamical patterns is critical to the function of those networked systems \cite{schnitzler2005normal,blaabjerg2006overview}.
For instance, long-range synchronization in the theta frequency band between the prefrontal cortex and the temporal cortex has been shown to improve working memory in older adults \cite{reinhart2019working}.

Up until now, synchronization (and other dynamical processes) in hypergraphs \cite{krawiecki2014chaotic,carletti2020dynamical,mulas2020coupled}, multilayer networks \cite{gambuzza2015intra,saa2018symmetries,belykh2019synchronization}, and temporal networks \cite{liberzon1999basic,belykh2004blinking,stilwell2006sufficient} have been studied mostly on a case-by-case basis.
Recently, it was shown that, in synchronization problems, simultaneous block diagonalization (SBD) optimally decouples the variational equation and enables the characterization of arbitrary synchronization patterns in large networks \cite{zhang2020symmetry}.
However, aside from multilayer networks, for which the multiple layers naturally translate into multiple matrices \cite{irving2012synchronization,zhang2020symmetry}, the full potential of SBD for analyzing dynamical patterns in generalized networks is yet to be realized.
As a technique, SBD has also found applications in numerous fields such as semi-definite programming \cite{maehara2011algorithm}, structural engineering \cite{murota1991computational}, signal processing \cite{cardoso1998multidimensional}, and quantum algorithms \cite{vspalek2008multiplicative}.

In this Article, we develop a versatile SBD-based framework that allows the stability analysis of synchronization patterns in generalized networks, which include hypergraphs, multilayer networks, and temporal networks.
This framework enables us to treat all three classes of generalized networks in a unified fashion. 
In particular, we show that different generalized interactions can all be represented by multiple matrices (as opposed to a single matrix as in the case of standard networks), and we introduce a practical method for finding the SBD of these matrices to simplify the stability analysis.
As an application of our unified framework, we use it to discover higher-order chimera states---intriguing cluster synchronization patterns that only emerge in the presence of nonpairwise couplings.

\vspace{-2mm}

\section{Results and Discussion}

\vspace{-3mm}

\subsection{General formulation and the SBD approach}

\vspace{-2mm}

Consider a general set of equations describing $N$ interacting oscillators:
\begin{equation}
  \mathbf{x}_i[t+1] = \mathbf{F}(\mathbf{x}_i[t]) + \mathbf{h}_i(\mathbf{x}_1[t],\cdots,\mathbf{x}_N[t],t),
\end{equation}
where $\mathbf{F}$ describes the intrinsic node dynamics and $\mathbf{h}_i$ specifies the influence of other nodes on node $i$.
We present our framework assuming discrete-time dynamics, although it works equally well for systems with continuous-time dynamics.

For a static network with a single type of pairwise interaction, $\mathbf{h}_i(\mathbf{x}_1,\cdots,\mathbf{x}_N,t) = \sigma\sum_{j=1}^{N} C_{ij} \mathbf{H}(\mathbf{x}_i,\mathbf{x}_j)$, where $\sigma$ is the coupling strength, the {\YZ (potentially weighted)} coupling matrix $\mathbf{C}$ reflects the network structure, and $\mathbf{H}$ is the interaction function.
When the network is globally synchronized, $\mathbf{x}_1=\cdots=\mathbf{x}_N=\mathbf{s}$, assuming $\mathbf{H}$ only depends on $\mathbf{x}_j$, the synchronization stability can be determined through the Lyapunov exponents associated with the variational equation
\begin{equation}
  \mathbf{\delta}[t+1] = \Big(\mathbf{I}_N \otimes \mathrm{J}\mathbf{F}(\mathbf{s}) + \sigma \mathbf{C} \otimes \mathrm{J}\mathbf{H}(\mathbf{s}) \Big) \mathbf{\delta}[t],
  \label{eq:vari_msf}
\end{equation}
where $\mathbf{\delta} = (\mathbf{x}_1^\intercal - \mathbf{s}^\intercal,\cdots, \mathbf{x}_N^\intercal - \mathbf{s}^\intercal)^\intercal$ is the perturbation vector, $\mathbf{I}_N$ is the identity matrix, $\otimes$ represents the Kronecker product, and $\mathrm{J}$ is the Jacobian operator.
In the case of undirected networks, \cref{eq:vari_msf} can always be decoupled into $N$ independent low-dimensional equations by switching to coordinates that diagonalize the coupling matrix $\mathbf{C}$ \cite{pecora1998master}.

For more complex synchronization patterns, however, additional matrices encoding information about dynamical clusters are inevitably introduced into the variational equation.
In particular, the identity matrix $\mathbf{I}_N$ is replaced by diagonal matrices $\mathbf{D}^{(m)}$ defined by
\begin{equation}
  D^{(m)}_{ii} = \begin{cases}
    1 & \text{if node } i \in \mathcal{C}_m, \\
    0 & \text{otherwise,}
  \end{cases}
\end{equation}
where $\mathcal{C}_m$ represents the $m$th dynamical cluster (oscillators within the same dynamical cluster are identically synchronized).
Moreover, as we show below, when $\mathbf{h}_i(\cdot)$ includes nonpairwise interactions, multilayer interactions, or time-varying interactions, it leads to additional coupling matrices $\mathbf{C}^{(k)}$ in the variational equation.
Consequently, the variational equations for complex synchronization patterns on generalized networks share the following form:
\begin{equation}
  \begin{split}
    \mathbf{\delta}[t+1]
    = & \bigg\{ \sum_{m} \mathbf{D}^{(m)} \otimes \mathrm{J}\mathbf{F}(\mathbf{s}^{m}) + \\
    & \sum_{m,k} \sigma_k\mathbf{C}^{(k)}\mathbf{D}^{(m)} \otimes \mathrm{J}\mathbf{H}^{(m,k)}(\mathbf{s}^m) \bigg\} \mathbf{\delta}[t],
  \label{eq:vari_general}
  \end{split}
\end{equation}
where $\mathbf{s}^{m}$ is the synchronized state of the oscillators in the $m$th dynamical cluster, and $\mathrm{J}\mathbf{H}^{(m,k)}(\mathbf{s}^m)$ is a Jacobian-like matrix whose expression depends on the class of generalized networks being considered.

For \cref{eq:vari_general}, diagonalizing any one of the matrices $\mathbf{D}^{(m)}$ or $\mathbf{C}^{(k)}$ generally does not lead to optimal decoupling of the equation.
Instead, all of the matrices $\mathbf{D}^{(m)}$ and $\mathbf{C}^{(k)}$ should be considered concurrently and be simultaneously block diagonalized to reveal independent perturbation modes. 
In particular, the new coordinates should separate the perturbation modes parallel to and transverse to the cluster synchronization manifold, and decouple transverse perturbations to the fullest extent possible.

For this purpose, we propose a practical algorithm to find an orthogonal transformation matrix $\mathbf{P}$ that simultaneously block diagonalizes multiple matrices.
Given a set of symmetric matrices $\mathcal{B} = \{\mathbf{B}^{(1)},\mathbf{B}^{(2)},\dots,\mathbf{B}^{(\mathscr{L})}\}$, the algorithm consists of three simple steps:
\begin{enumerate}[i.]
 \item Find the (orthogonal) eigenvectors $\mathbf{v}_i$ of the matrix $\mathbf{B} = \sum_{\ell=1}^\mathscr{L} \xi_\ell \mathbf{B}^{(\ell)}$, where $\xi_\ell$ are independent random coefficients {\YZ which can be} drawn from a Gaussian distribution. Set $\mathbf{Q} = [\mathbf{v}_1,\cdots,\mathbf{v}_N]$.
 \item Generate $\mathbf{B} = \sum_{\ell=1}^\mathscr{L} \xi_\ell \mathbf{B}^{(\ell)}$ for a new realization of $\xi_\ell$ and compute $\widetilde{\mathbf{B}}=\mathbf{Q}^\intercal \mathbf{B} \mathbf{Q}$. Mark the indexes $i$ and $j$ as being in the same block if $\widetilde{B}_{ij}\neq 0$ (and thus $\widetilde{B}_{ji}\neq 0$).
 \item Set $\mathbf{P} = [\mathbf{v}_{\epsilon(1)},\cdots,\mathbf{v}_{\epsilon(N)}]$, where $\epsilon$ is a permutation of $1,\cdots,N$ such that indexes in the same block are sorted consecutively (i.e., the base vectors $\mathbf{v}_i$ corresponding to the same block are grouped together).
\end{enumerate}

The proposed algorithm is inspired by and adapted from 
Murota et al.\ \cite{murota2010numerical}.
There, the authors use the eigendecompostion of a random linear combination of the given matrices to find a partial SBD, but the operations needed for refining the blocks can be cumbersome.
Here, we show that the simplified algorithm above is guaranteed to find the finest SBD when there are no degeneracies---i.e., no two $\mathbf{v}_i$ have the same eigenvalue (see Methods for a proof).
Intuitively, this is because a random linear combination of $\mathbf{B}^{(\ell)}$ contains all the information about their common block structure in the absence of degeneracy, which can be efficiently extracted through eigendecompostion.
When there is a degeneracy, cases exist for which the proposed algorithm does not find the finest SBD (see Methods for details).
However, these cases are rare in practice and is a small price to pay for the improved simplicity and efficiency of the algorithm.

We note that the algorithm can be adapted to asymmetric matrices, and in all nondegenerate cases it finds the finest SBD that can be achieved by orthogonal transformations. 
However, this does not exclude the possibility that more general similarity transformations could result in finer blocks for certain asymmetric matrices.
(For symmetric matrices, general similarity transformations do not have an advantage over orthogonal transformations.)

\begin{figure}[tb]
\centering
\includegraphics[width=\columnwidth]{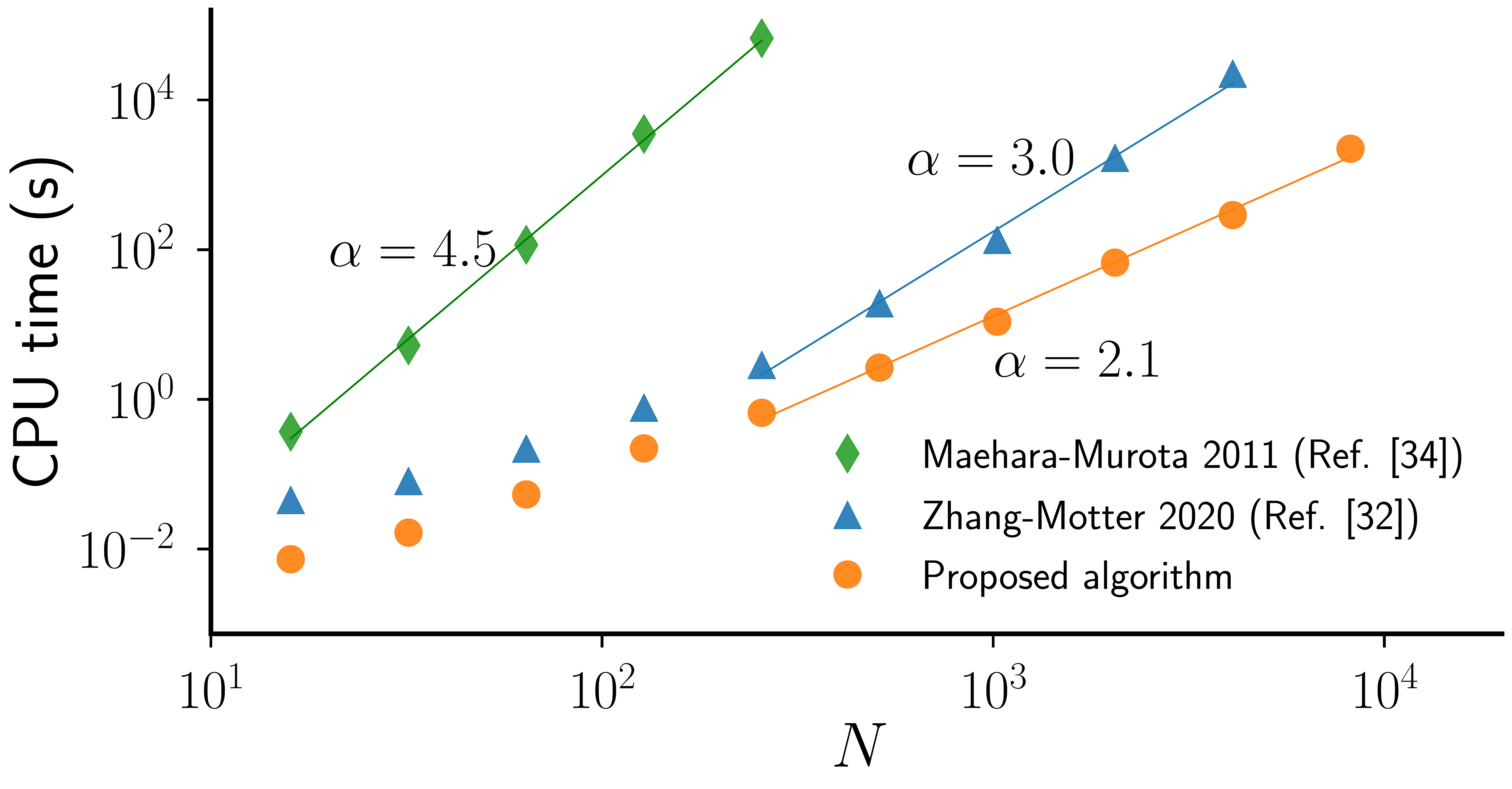}
\vspace{-4mm}
\caption{{\bf Computational costs of different simultaneous block diagonalization algorithms as functions of matrix size $N$.} 
The computational costs of all three algorithms scale as $N^\alpha$ for large $N$. However, the algorithm proposed here has the smallest exponent $\alpha$, which translates to order-of-magnitude speedups already for moderate matrix sizes.
}
\label{fig:sbd}
\end{figure}

In \cref{fig:sbd}, we compare the proposed algorithm with two previous state-of-the-art algorithms on SBD \cite{maehara2011algorithm,zhang2020symmetry}.
The algorithms are tested on sets of $N\times N$ matrices, each consisting of $10$ random matrices with predefined common block structures (see Methods for how the matrices are generated).
For each algorithm and each matrix size $N$, $100$ independent matrix sets are tested.
\Cref{fig:sbd} shows the mean CPU time from each set of tests (the standard deviations are smaller than the size of the symbols).
The algorithm presented here finds the finest SBD in all cases tested and has the most favorable scaling in terms of computational complexity.
For instance, it can process matrices with $N\approx 1000$ in under $10$ seconds (tested on Intel Xeon E5-2680v3 processors), which is orders of magnitude faster than the other methods.
The Python and MATLAB implementations of the proposed SBD algorithm are available online as part of this publication (see Code availability).

\subsection{Cluster synchronization and chimera states in hypergraphs}

Hypergraphs \cite{berge1973graphs} and simplicial complexes \cite{hatcher2002algebraic} provide a general description of networks with nonpairwise interactions and have been widely adopted in the literature \cite{petri2014homological,giusti2016two,benson2016higher,bairey2016high,mayfield2017higher,levine2017beyond,patania2017shape,reimann2017cliques,sizemore2018cliques,benson2018simplicial,petri2018simplicial,kuzmin2018systematic,tekin2018prevalence,estrada2018centralities,iacopini2019simplicial,leon2019phase,matheny2019exotic,matamalas2020abrupt,de2020social,schaub2020random,carletti2020random,landry2020effect,st2021master}.
However, the associated tensors describing those higher-order structures are more involved than matrices, especially when combined with the analysis of dynamical processes \cite{tanaka2011multistable,bick2016chaos,skardal2019abrupt,skardal2020higher,xu2020bifurcation,millan2020explosive}.
There have been several efforts to generalize the master stability function (MSF) formalism \cite{pecora1998master} to these settings, for which different variants of an aggregated Laplacian have been proposed \cite{mulas2020coupled,lucas2020multiorder,carletti2020dynamical,de2021phase}.
The aggregated Laplacian captures interactions of all orders in a single matrix, whose spectral decomposition allows the stability analysis to be divided into structural and dynamical components, just like the standard MSF for pairwise interactions.
However, such powerful reduction comes at an inevitable cost: simplifying assumptions must be made about the network structure (e.g., all-to-all coupling), node dynamics (e.g., fixed points), and/or interaction functions (e.g., linear) in order for the aggregation to a single matrix to be valid.

Here, we consider general oscillators coupled on hypergraphs without the aforementioned restrictions. 
For the ease of presentation and without loss of generality, we focus on networks with interactions that involve up to three oscillators simultaneously:
\begin{equation}
	\begin{aligned}
		\mathbf{x}_i[t+1] = & \mathbf{F}\left(\mathbf{x}_i[t]\right) + \sigma_{1} \sum_{j=1}^{N} A_{ij}^{(1)} \mathbf{H}^{(1)}\left(\mathbf{x}_i[t],\mathbf{x}_j[t]\right) \\
		+ & \sigma_{2} \sum_{j=1}^{N} \sum_{k=1}^{N} A_{ijk}^{(2)} \mathbf{H}^{(2)}\left(\mathbf{x}_i[t],\mathbf{x}_j[t],\mathbf{x}_k[t]\right).
    \label{eq:dyn_hyper}
	\end{aligned}
\end{equation}
The adjacency matrix $\mathbf{A}^{(1)}$ and adjacency tensor $\mathbf{A}^{(2)}$ represent the pairwise and the three-body interaction, respectively.
To make progress, we use the following key insight from Gambuzza et al.\ \cite{gambuzza2021stability}:
for noninvasive coupling [i.e., $\mathbf{H}^{(1)}(\mathbf{s},\mathbf{s}) = 0$ and $\mathbf{H}^{(2)}(\mathbf{s},\mathbf{s},\mathbf{s}) = 0$] and global synchronization, synchronization stability in hypergraphs is determined by \cref{eq:vari_general} with $\mathbf{C}^{(k)} = -\mathbf{L}^{(k)}$, where $\mathbf{L}^{(k)}$ are generalized Laplacians defined based on the adjacency tensors $\mathbf{A}^{(k)}$.
More concretely, $\mathbf{L}^{(1)}$ is the usual Laplacian, for which $L_{ij}^{(1)} = \delta_{ij} \sum_k A_{ik}^{(1)} - A_{ij}^{(1)}$; $\mathbf{L}^{(2)}$ retains the zero row-sum property and is defined as $L^{(2)}_{ij} = -\sum_k A^{(2)}_{ijk}$ for $i \neq j$ and $L^{(2)}_{ii} = -\sum_{k\neq i} L^{(2)}_{ik}$.
Higher-order generalized Laplacians for $k > 2$ can be defined similarly \cite{gambuzza2021stability}.

\begin{figure*}[tb]
\centering
\includegraphics[width=.7\textwidth]{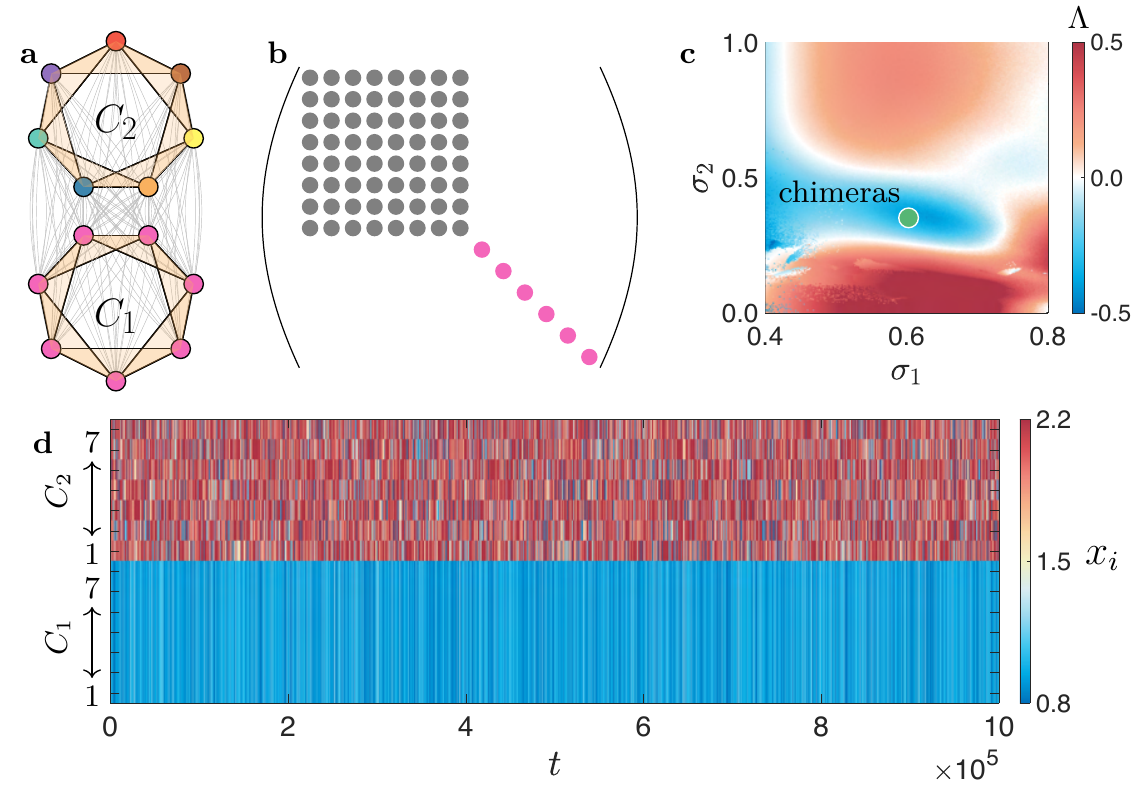}
\vspace{-4mm}
\caption{{\bf Chimera states arising from nonpairwise interactions.}
{\bf a} Two identical subnetworks ($C_1$ and $C_2$) of optoelectronic oscillators with strong intracluster connections (black lines) and weak intercluster connections (gray lines). The three-body interactions are indicated by 2-simplices (beige triangles). The eight dynamical clusters that form the chimera state are indicated by different node colors.
{\bf b} Common block structure of the matrices in the variational equation (\ref{eq:vari_general}) revealed by the SBD algorithm, in which nonzero entries are represented by solid circles. The gray block corresponds to perturbations parallel to the synchronization manifold, and the pink blocks represent perturbations transverse to the synchronization manifold. Thus, only the pink blocks need to be considered in the stability analysis. For the network in {\bf a}, the transverse perturbations are all localized within the subnetwork $C_1$.
{\bf c} Linear stability analysis of chimera states based on the SBD coordinates for a range of the pairwise interaction strength $\sigma_1$ and three-body interaction strength $\sigma_2$. Chimeras are stable when the maximum transverse Lyapunov exponent $\Lambda$ is negative, and they occur only in the presence of nonvanishing three-body interactions.
{\bf d} Chimera dynamics for $\sigma_1=0.6$ and $\sigma_2=0.4$ (green dot in {\bf c}). Here, $x_i$ is the dynamical state of the $i$th oscillator, and the vertical axis indexes the oscillators in the respective subnetworks.
}
\label{fig:chimera_hyper}
\end{figure*}

Crucially, we can show that the generalized Laplacians are sufficient for the stability analysis of cluster synchronization patterns provided that the clusters are nonintertwined \cite{pecora2014cluster,cho2017stable} (see Supplementary Note 1 for a mathematical derivation).
Thus, in these cases, the problem reduces to applying the SBD algorithm to the set formed by matrices $\{\mathbf{D}^{(m)}\}$ (determined by the synchronization pattern) and $\{\mathbf{L}^{(k)}\}$ (encoding the hypergraph structure).
For the most general case that includes intertwined clusters, SBD still provides the optimal reduction, as long as the generalized Laplacians are replaced by matrices that encode more nuanced information about the relation between different clusters \cite{salova2021cluster}.
The resulting SBD coordinates significantly simplifies the calculation of Lyapunov exponents in \cref{eq:vari_general} and can provide valuable insight on the origin of instability, as we show below.

As an application to nontrivial synchronization patterns, we study chimera states \cite{panaggio2015chimera,omel2018mathematics} on hypergraphs.
Here, chimera states are defined as spatiotemporal patterns that emerge in systems of identically coupled identical oscillators in which part of the oscillators are mutually synchronized while the others are desynchronized.
For a comprehensive review on different notions of chimeras, see the recent survey by Haugland \cite{haugland2021changing}.

The hypergraph in \cref{fig:chimera_hyper}a consists of two subnetworks of optoelectronic oscillators. 
Each subnetwork is a simplicial complex, in which a node is coupled to its four nearest neighbors through pairwise interactions of strength $\sigma_1$ and it also participates in three-body interactions of strength $\sigma_2$.
The two subnetworks are all-to-all coupled through weaker links of strength $\kappa\sigma_1$, and in our simulations we take $\kappa=1/5$. 
The individual oscillators are modeled as discrete maps $x_i[t+1] = \beta\sin^2\big(x_i[t]+\pi/4\big)$, where $\beta$ is the self-feedback strength that is tunable in experiments \cite{hart2017experiments,hart2019topological}.
For the pairwise interaction, we set $H^{(1)}(x_i,x_j) = \sin^2\big(x_j+\pi/4\big) - \sin^2\big(x_i+\pi/4\big)$.
For the three-body interaction, we set $H^{(2)}(x_i,x_j,x_k) = \sin^2\big(x_j+x_k-2x_i\big)$.
The full dynamical equation of the system can be summarized as follows:
\begin{equation}
  \begin{split}
    x_i[t+1] = & \,\beta\,\sin^2\big(x_i[t]+\frac{\pi}{4}\big) \\
    + & \sigma_1 \sum_{j=1}^N A^{(1)}_{ij} \left( \sin^2\big(x_j[t]+\frac{\pi}{4}\big) - \sin^2\big(x_i[t]+\frac{\pi}{4}\big)\right) \\ 
    + & \sigma_2 \sum_{j=1}^N\sum_{k=1}^N A^{(2)}_{ijk} \sin^2\big(x_j[t]+x_k[t]-2x_i[t]\big).
\end{split}
\label{eq:chimera}
\end{equation}
Since couplings in previous optoelectronic experiments are implemented through a field-programmable gate array that can realize three-body interactions, we expect that our predictions below can be explored and verified experimentally on the same platform.

To characterize chimera states for which one subnetwork is synchronized and one subnetwork is incoherent, we are confronted with $10$ noncommuting matrices in \cref{eq:vari_general}.
Eight of them are $\{\mathbf{D}^{(1)},\cdots,\mathbf{D}^{(8)}\}$, corresponding to one dynamical cluster with $7$ synchronized nodes and seven dynamical clusters with $1$ node each (distinguished by colors in \cref{fig:chimera_hyper}a).
The other two matrices are $\{\mathbf{L}^{(1)},\mathbf{L}^{(2)}\}$, which describe the pairwise and three-body interactions, respectively.
Applying the SBD algorithm to these matrices reveals the common block structure depicted in \cref{fig:chimera_hyper}b.
The gray block corresponds to perturbations parallel to the cluster synchronization manifold and does not affect the chimera stability.
The other blocks control the transverse perturbations (all localized within the synchronized subnetwork $C_1$) and are included in the stability analysis.
This allows us to focus on one $1\times 1$ block at a time and to efficiently calculate the maximum transverse Lyapunov exponent (MTLE) $\Lambda$ of the chimera state using previously established procedure \cite{zhang2020critical,zhang2021mechanism}.

\begin{figure*}[t]
\includegraphics[width=1.4\columnwidth]{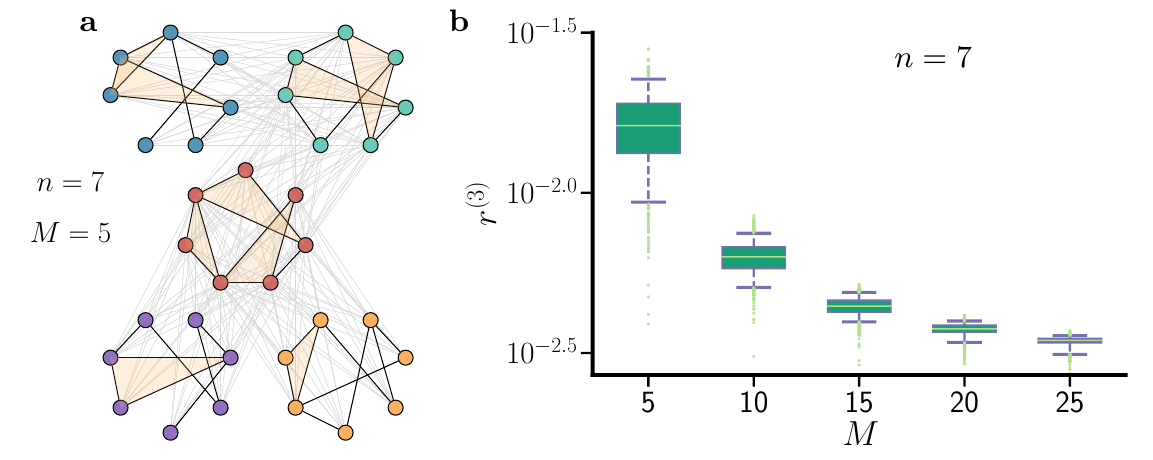}
\caption{{\bf Complexity reduction in the analysis of synchronization patterns in hypergraphs.}
{\bf a} Example of a hypergraph consisting of $M=5$ clusters, each with $n=7$ nodes. Inside each cluster there are pairwise interactions (black lines) and three-body interactions (beige triangles). Two clusters are either all-to-all connected (gray lines) or fully disconnected.
{\bf b} Reduction in computational complexity achieved by the SBD algorithm for cluster size $n=7$, intracluster link density $p=0.5$, and intercluster link density $q=0.5$ as the cluster number $M$ is varied. The box covers the range $25^{\text{th}}$--$75^{\text{th}}$ percentile, the whiskers mark the range $5^{\text{th}}$--$95^{\text{th}}$ percentile, and the dots indicate the remaining $10\%$ outliers. Each boxplot is based on $1000$ independent network realizations.
}
\label{fig:sbd_hyper}
\end{figure*}

For the system in \cref{fig:chimera_hyper}, SBD coordinates offer not only dimension reduction but also analytical insights. 
As we show in Supplementary Note 2, because the transverse blocks (colored pink in \cref{fig:chimera_hyper}b) found by the SBD algorithm are all $1\times 1$, the Lyapunov exponents associated with chimera stability are given by a simple formula, 
\begin{equation}
  \Lambda_i = \ln \left|1 - \frac{\sigma_1}{\beta} \left(\lambda_i^{(1)} + \frac{\kappa N}{2} \right) \right| + \Gamma,
  \label{eq:chimera_stability}
\end{equation}
where $\lambda_i^{(1)}$ is the scalar inside the $i$th transverse block of $\mathbf{L}^{(1)}$ after the SBD transformation.
Here,
\begin{equation}
  \begin{aligned}
    \Gamma = & \lim_{\mathcal{T}\rightarrow \infty} \frac{1}{\mathcal{T}} \sum_{t=1}^{\mathcal{T}} \ln \left| \mathrm{J}F(s[t]) \right| \\
    = & \lim_{\mathcal{T}\rightarrow \infty} \frac{\beta}{\mathcal{T}} \sum_{t=1}^{\mathcal{T}} \ln \left| \sin\big(2s[t]+\frac{\pi}{2}\big) \right|
  \end{aligned}
  \label{eq:gamma}
\end{equation}
is a finite constant determined by the synchronous trajectory $s[t]$ of the coherent subnetwork $C_1$, which in turn is influenced by both $\sigma_1$ and $\sigma_2$.

Using \cref{eq:chimera_stability,eq:gamma}, we can calculate the MTLE in the $\sigma_1$-$\sigma_2$ parameter space to map out the stable chimera region.
As can be seen from \cref{fig:chimera_hyper}c, where we fix $\beta=1.5$, chimera states are unstable when oscillators are coupled only through pairwise interactions (i.e., when $\sigma_2=0$),
but they become stable in the presence of three-body interactions of intermediate strength.
\Cref{fig:chimera_hyper}d shows the typical chimera dynamics for $\beta = 1.5$, $\sigma_1 = 0.6$, and $\sigma_2 = 0.4$.
According to \cref{eq:chimera_stability,eq:gamma}, the higher-order interaction stabilizes chimera states solely by changing the dynamics in the incoherent subnetwork $C_2$, which in turn influences the synchronous trajectory in $C_1$ and thus the value of $\Gamma$.
This insight highlights the critical role played by the incoherent subnetwork in determining chimera stability \cite{zhang2021mechanism}.

To test the complexity reduction capability of the SBD algorithm systematically, we consider networks consisting of $M$ dynamical clusters, each with $n$ nodes (\cref{fig:sbd_hyper}a), such that:
\begin{enumerate}
  \item each cluster is a random subnetwork with link density $p$, to which three-body interactions are added by transforming triangles into 2-simplices;
  \item two clusters are either all-to-all connected (with probability $q>0$) or fully disconnected from each other (with probability $1-q$).
\end{enumerate}
For the analysis of the $M$-cluster synchronization state in these networks, the reduction in computational complexity yielded by the SBD algorithm can be measured
using $r^{(\alpha)} = \sum_i n_i^\alpha/N^\alpha$, where $n_i$ is the size of the $i$th common block for the transformed matrices.
If the computational complexity of analyzing \cref{eq:vari_general} in its original form scales as $\mathcal{O}(N^\alpha)$, then $r^{(\alpha)}$ gives the fraction of time needed to analyze \cref{eq:vari_general} in its decoupled form under the SBD coordinates.
Given that the computational complexity of finding the Lyapunov exponents for a fixed point in an $n$-dimensional space typically lies between $\mathcal{O}(n^2)$ and $\mathcal{O}(n^3)$, here we set $\alpha=3$ as a reference for the more challenging task of calculating the Lyapunov exponents for periodic or chaotic trajectories.

In \cref{fig:sbd_hyper}b, we apply the SBD algorithm to $\{\mathbf{D}^{(1)},\cdots,\mathbf{D}^{(M)},\mathbf{L}^{(1)},\mathbf{L}^{(2)}\}$ and plot $r^{(3)}$ against the number of clusters $M$ in the networks.
We see a reduction in complexity of at least two orders of magnitude ($r^{(3)} \leq 10^{-2}$) for $M\geq 10$.
This reduction does not depend sensitively on other parameters in our model ($n$, $p$, and $q$).

\begin{figure*}[tb]
\centering
\includegraphics[width=.7\textwidth]{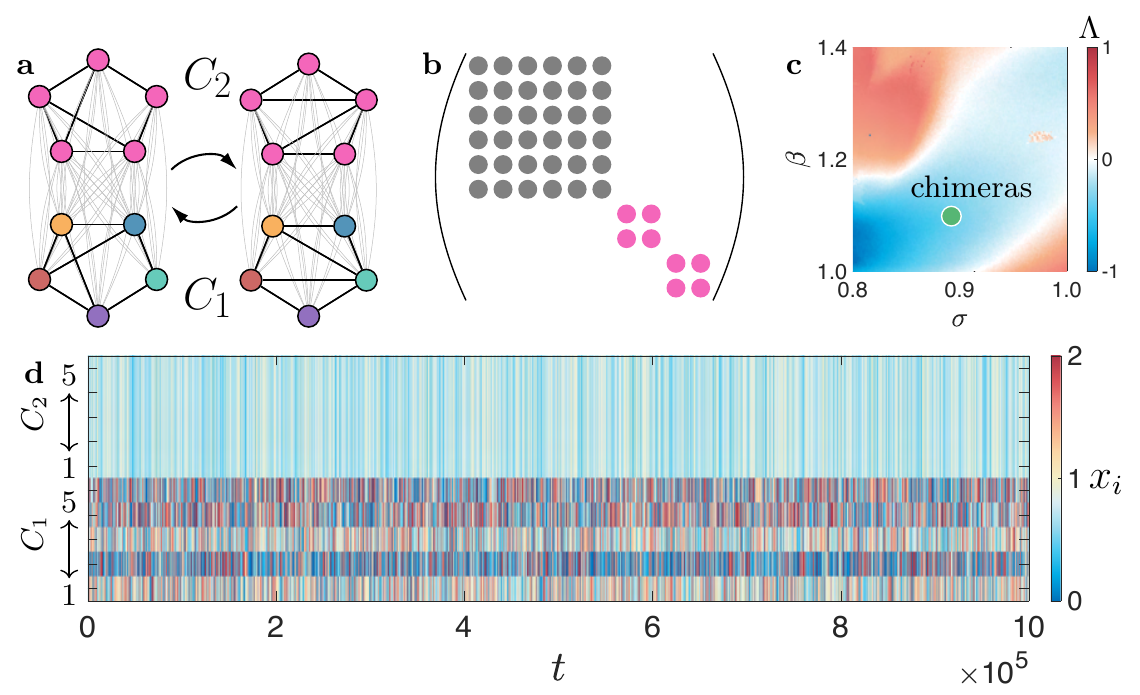}
\vspace{-4mm}
\caption{{\bf Chimera states on a temporal network.} 
{\bf a} Two identical subnetworks of optoelectronic oscillators with strong intracluster connections (black lines) and weak intercluster connections (gray lines). The network structure switches back and forth between two different configurations. The six dynamical clusters that form the chimera state are indicated by different node colors.
{\bf b} Common block structure of the matrices in the variational equation (\ref{eq:vari_general}) under the SBD coordinates. The entries of the transformed matrices that are not required to be zero are represented by solid circles. The gray block corresponds to perturbations that do not affect the chimera stability, and the pink blocks represent transverse perturbations that determine the chimera stability.
{\bf c} Linear stability analysis of chimera states based on the SBD coordinates for a range of coupling strength $\sigma$ and self-feedback strength $\beta$. Chimeras are stable when the maximum transverse Lyapunov exponent $\Lambda$ is negative.
{\bf d} Chimera dynamics for $\sigma=0.9$ and $\beta=1.1$ (green dot in {\bf c}). Here, $x_i$ is the dynamical state of the $i$th oscillator, and the vertical axis indexes the oscillators in the respective subnetworks marked in {\bf a}.
}
\label{fig:chimera_temporal}
\end{figure*}

\subsection{Synchronization patterns in multilayer and temporal networks}

The coexistence of different types (i.e., layers) of interactions in a network \cite{kivela2014multilayer,boccaletti2014structure,aleta2019multilayer} can dramatically influence underlying dynamical processes, such as percolation \cite{cellai2013percolation,osat2017optimal}, diffusion \cite{gomez2013diffusion,de2016physics}, and synchronization \cite{jalan2016cluster,nicosia2017collective,belykh2019synchronization}.
Multilayer networks of $N$ oscillators diffusively coupled through $K$ different types of interactions can be described by
\begin{equation}
  \mathbf{x}_i[t+1] = \mathbf{F}(\mathbf{x}_i[t]) - \sum_{k=1}^{K} \sigma_k \sum_{j=1}^{N} L_{ij}^{(k)} \mathbf{H}^{(k)}(\mathbf{x}_j[t]),
\end{equation}
where $\mathbf{L}^{(k)}$ is the Laplacian matrix representing the links mediating interactions of the form $\mathbf{H}^{(k)}$ and coupling strength $\sigma_k$.
It is easy to see that the corresponding variational equation for a given synchronization pattern \cite{zhang2020symmetry,della2020symmetries} is a special case of \cref{eq:vari_general} and can be readily addressed using the SBD framework.

\begin{figure*}[tbh!]
\includegraphics[width=1.4\columnwidth]{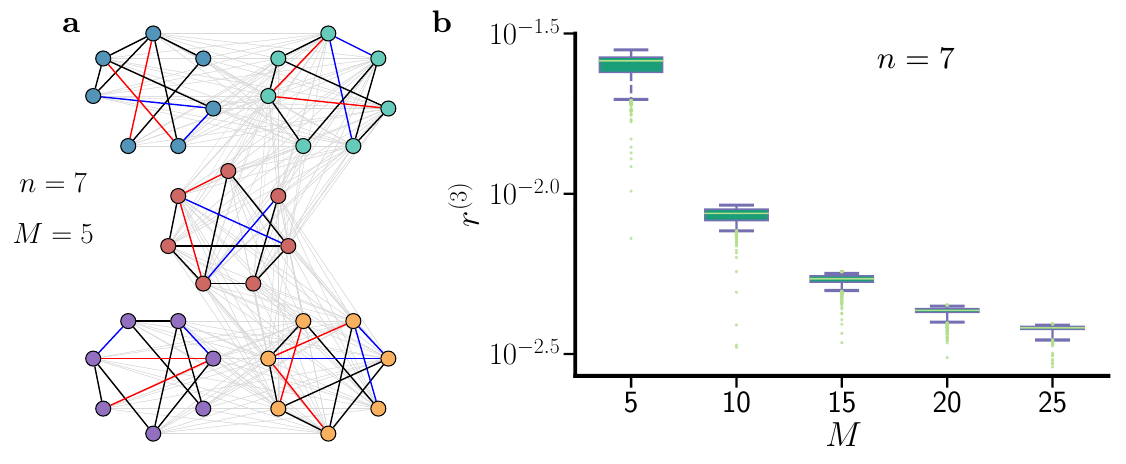}
\caption{{\bf Complexity reduction in the analysis of synchronization patterns in temporal networks.} 
{\bf a} Example of a temporal network consisting of $M=5$ clusters, each with $n=7$ nodes. In each cluster, an expected 20\% of the links are temporal (connections alternate between the blue and the red links) and the remaining 80\% are static (black links). Two clusters are either all-to-all connected (gray lines) or fully disconnected.
{\bf b} Reduction in computational complexity achieved by the SBD algorithm for cluster size $n=7$, intracluster link density $p=0.5$, and intercluster link density $q=0.5$ as the cluster number $M$ is varied. The box covers the range $25^{\text{th}}$--$75^{\text{th}}$ percentile, the whiskers mark the range $5^{\text{th}}$--$95^{\text{th}}$ percentile, and the dots indicate the remaining $10\%$ outliers. Each boxplot is based on $1000$ independent network realizations.
}
\label{fig:sbd_temporal}
\end{figure*}

Temporal networks \cite{holme2012temporal} are another class of systems that can naturally be addressed using our SBD framework.  
Such networks are ubiquitous in nature and society \cite{li2017fundamental,paranjape2017motifs}, and their time-varying nature has been shown to significantly alter many dynamical characteristics, including controllability \cite{posfai2014structural,li2017fundamental} and synchronizability \cite{amritkar2006synchronized,lu2008synchronization,jeter2015synchronization,zhang2021designing}.

Consider a temporal network whose connection pattern at time $t$ is described by $\mathbf{L}^{(t)}$,
\begin{equation}
  \mathbf{x}_i[t+1] = \mathbf{F}(\mathbf{x}_i[t]) - \sigma \sum_{j=1}^{N} L_{ij}^{(t)} \mathbf{H}(\mathbf{x}_j[t]).
\end{equation}
Here, the stability analysis of synchronization patterns can by simplified by simultaneously block diagonalizing $\{\mathbf{D}^{(m)}\}$ and $\{\mathbf{L}^{(t)}\}$.
This framework generalizes existing master stability methods for synchronization in temporal networks \cite{boccaletti2006synchronization},
which assumes that synchronization is global and the set of all $\mathbf{L}^{(t)}$ to be commutative.
We also do not require separation of time scales between the evolution of the network structure and the internal dynamics of oscillators, which was assumed in various previous studies in exchange of analytical insights \cite{belykh2004blinking,stilwell2006sufficient}.
It is worth noting that $\{\mathbf{L}^{(t)}\}$ can in principle contain infinitely many different matrices.
This would pose a challenge to the SBD algorithm unless there are relations among the matrices to be exploited.
Here, for simplicity, we assume that $\mathbf{L}^{(t)}$ are selected from a finite set of matrices.
{\YZ This class of temporal networks is also referred to as switched systems in the engineering literature and has been widely studied \cite{liberzon1999basic}.}

As an application, we characterize chimera states on a temporal network that alternates between two different configurations.
\Cref{fig:chimera_temporal}a illustrates the temporal evolution of the network, which has intracluster coupling of strength $\sigma$ and intercluster coupling of strength $\kappa\sigma$ (again for $\kappa=1/5$, the same optoelectronic oscillator and pairwise interaction function as in \cref{fig:chimera_hyper}).
This system has a variational equation with noncommuting matrices $\{\mathbf{D}^{(1)}, \cdots, \mathbf{D}^{(6)}, \mathbf{L}^{(1)}, \mathbf{L}^{(2)}\}$, where $\mathbf{L}^{(1)}$ and $\mathbf{L}^{(2)}$ correspond to the network configuration at odd and even $t$, respectively.
Applying the SBD algorithm reveals one $6\times 6$ parallel block and two $2\times 2$ transverse blocks (\cref{fig:chimera_temporal}b), effectively reducing the dimension of the stability analysis problem from $10$ to $2$.

Despite the transverse blocks not being $1\times 1$, by looking at the transformation matrix $\mathbf{P}$ one can still gather insights about the nature of the instability.
For example, the first pink block consists of transverse perturbations (localized in the synchronized subnetwork) of the form $\big(a,0,-a,b,-b\big)$, while perturbations in the second pink block are constrained to be $\big(c,-2(c+d),c,d,d\big)$.
Depending on which block becomes unstable first, the synchronized subnetwork (and thus the chimera state) loses stability through different routes.
The chimera region based on the MTLE calculated under the SBD coordinates is shown in \cref{fig:chimera_temporal}c and the typical chimera dynamics for $\sigma=0.9$ and $\beta=1.1$ are presented in \cref{fig:chimera_temporal}d.

To further demonstrate the utility of the SBD framework, we systematically consider temporal networks that alternate between two different configurations.
The network construction is similar to that in \cref{fig:sbd_hyper}, except that here each cluster has time-varying instead of nonpairwise interactions.
In the example shown in \cref{fig:sbd_temporal}a, each cluster has red links active at odd $t$ and blue links active at even $t$, while the black links are always active.
\Cref{fig:sbd_temporal}b confirms that the SBD algorithm consistently leads to substantial reduction in computational complexity. 
Moreover, as in the case of hypergraphs (\cref{fig:sbd_hyper}), the complexity reduction increases as the number of clusters $M$ is increased.
Again, the results do not depend sensitively on cluster size and link densities.

\section{Conclusion}

In this work, we established SBD as a versatile tool to analyze complex synchronization patterns in generalized networks with nonpairwise, multilayer, and time-varying interactions.
The method can be easily applied to other dynamical processes, such as diffusion \cite{de2016physics}, random walks \cite{schaub2020random}, and consensus \cite{neuhauser2020multibody}.
Indeed, the equations describing such processes on generalized networks often involve two or more noncommuting matrices, whose SBD naturally leads to an optimal mode decoupling and the simplification of the analysis.

The usefulness of our framework also extends beyond the generalized networks discussed here. 
Many real-world networks are composed of different types of nodes and can experience nonidentical delays in the communications among nodes.
These heterogeneities can be represented through additional matrices and are automatically accounted for by our SBD framework in the stability analysis \cite{zhang2017nonlinearity}.
Finally, we suggest that our results may find applications beyond network dynamics, since SBD is also a powerful tool to address other problems involving multiple matrices in which dimension reduction is desired, such as independent component analysis and blind source separation \cite{choi2005blind,comon2010handbook}.
The flexibility and scalability of our framework make it adaptable to various practical situations, and we thus expect it to facilitate the exploration of collective dynamics in a broad range of complex systems.

\section{Methods}

\begin{figure}[t]
\centering
\includegraphics[width=.8\columnwidth]{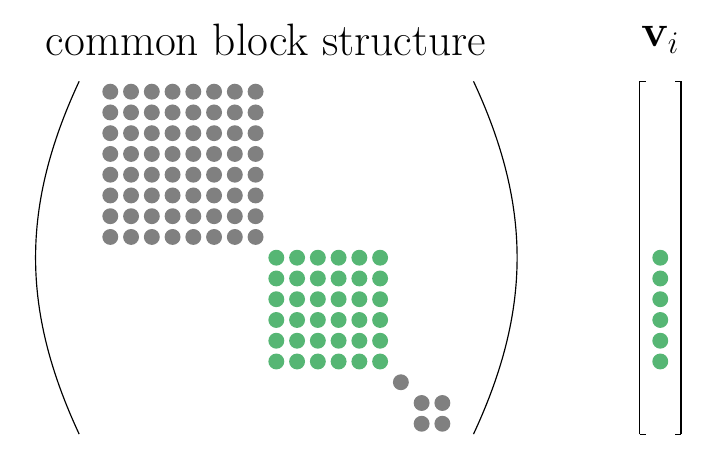}
\caption{{\bf Illustration of a localized eigenvector.} The vector $\mathbf{v}_i$ is localized within the green block of the matrix. The nonzero entries of the matrix and the vector are represented as solid circles.}
\label{fig:6}
\end{figure}

\noindent\textbf{Optimality of the common block structure discovered by the SBD algorithm.}
Given a set of symmetric matrices $\mathcal{B} = \{\mathbf{B}^{(1)},\mathbf{B}^{(2)},\dots,\mathbf{B}^{(\mathscr{L})}\}$, let $\mathbf{B} = \sum_{\ell=1}^{\mathscr{L}} \xi_\ell \mathbf{B}^{(\ell)}$, where $\xi_\ell$ are random coefficients.
Without loss of generality, we can assume all matrices $\mathbf{B}^{(\ell)}$ to be in their finest common block form.
Our goal is then to prove that, when there is no degeneracy, each eigenvector $\mathbf{v}_i$ of $\mathbf{B}$ is localized within a single (square) block, meaning that the indices of the nonzero entries of $\mathbf{v}_i$ are limited to the rows of one of the common blocks shared by $\{\mathbf{B}^{(\ell)}\}$ (\cref{fig:6}).

We first notice that $\mathbf{B}$ inherits the common block structure of $\{\mathbf{B}^{(\ell)}\}$.
Thus, for each $n_i\times n_i$ block shared by $\{\mathbf{B}^{(\ell)}\}$, we can always find $n_i$ eigenvectors of $\mathbf{B}$ that are localized within that block. 
When the eigenvalues of $\mathbf{B}$ are nondegenerate, the eigenvectors are unique, and thus all $N=\sum_i n_i$ eigenvectors of matrix $\mathbf{B}$ are localized within individual blocks.

Based on the results above, it follows that after computing the eigenvectors $\mathbf{v}_i$ of matrix $\mathbf{B}$ (step i of the SBD algorithm) and sorting them according to their associated block (steps ii and iii of the SBD algorithm), the resulting orthogonal matrix $\mathbf{P} = [\mathbf{v}_{\epsilon(1)},\cdots,\mathbf{v}_{\epsilon(N)}]$ will reveal the finest common block structure.
Here, {\it finest} is characterized by the number of common blocks being maximal (which is also equivalent to the sizes of the blocks being minimal), and the block sizes are unique up to permutations.

In the presence of degeneracies (i.e., when there are distinct eigenvectors with the same eigenvalue), no theoretical guarantee can be given that the strategy above will find the finest SBD \cite{murota2010numerical}. 
To see why, consider the matrices $\mathbf{B}^{(\ell)}=\text{diag}(\mathbf{b}^{(\ell)},\mathbf{b}^{(\ell)},\dots,\mathbf{b}^{(\ell)})$ formed by the direct sum of duplicate blocks.
In this case, a generic $\mathbf{B}$ has eigenvalues with multiplicity $M$, where $M$ is the number of duplicate blocks.
For example, if $\mathbf{u}$ is an eigenvector corresponding to the first block of $\mathbf{B}$, then $(\xi_1\mathbf{u}^\intercal,\dots,\xi_M\mathbf{u}^\intercal)^\intercal$ is also an eigenvector of $\mathbf{B}$ (with the same eigenvalue) for any set of random coefficients $\{\xi_m\}$.
As a result, the eigenvectors of $\mathbf{B}$ are no longer guaranteed to be localized within a single block.

\vspace{3mm}

\noindent\textbf{Generating random matrices with predefined block structures.}
In order to compare the computational costs of different SBD algorithms, we generate sets of random matrices with predefined common block structures.
For each set, we start with $\mathscr{L}=10$ matrices of size $N$.
The $\ell$th matrix is constructed as the direct sum of smaller random matrices, $\mathbf{B}^{(\ell)}=\text{diag}(\mathbf{b}_1^{(\ell)},\dots,\mathbf{b}_M^{(\ell)})$, where $\mathbf{b}_m^{(\ell)}$ are symmetric matrices with entries drawn from the Gaussian distribution $\mathcal{N}(0,1)$.
The size of the $m$th block $\mathbf{b}_m^{(\ell)}$ is chosen randomly between $1$ and $N/2$ and is set to be the same for all $\ell$.
We then apply a random orthogonal transformation $\mathbf{Q}$ to $\mathcal{B} = \{\mathbf{B}^{(1)},\mathbf{B}^{(2)},\dots,\mathbf{B}^{(\mathscr{L})}\}$, which results in a matrix set $\widetilde{\mathcal{B}} = \{\widetilde{\mathbf{B}}^{(1)},\widetilde{\mathbf{B}}^{(2)},\dots,\widetilde{\mathbf{B}}^{(\mathscr{L})}\}$ with no apparent block structure in $\widetilde{\mathbf{B}}^{(\ell)}=\mathbf{Q}^\intercal \mathbf{B}^{(\ell)} \mathbf{Q}$.
Finally, the SBD algorithms are applied to $\widetilde{\mathcal{B}}$ to recover the common block structure.
All tests are performed on Intel Xeon E5-2680 v3 Processors, and the CPU time used by each algorithm is recorded using the {\texttt timeit} function from MATLAB.


\clearpage

\noindent \textbf{Supplementary information} is available for this paper.\\
\noindent \textbf{Acknowledgements:} The authors thank Fiona Brady, Takanori Maehara, Anastasiya Salova, and Raissa D'Souza for insightful discussions. This work was supported by the  U.S. Army Research Office (Grant No.\ W911NF-19-1-0383). Y.Z. was further supported by a Schmidt Science Fellowship. V.L. acknowledges support from the Leverhulme Trust Research Fellowship ``CREATE: The Network Components of Creativity and Success'' and the Engineering and Physical Sciences Research Council (Grant No.\ EP/N013492/1).\\
\noindent \textbf{Author contributions:} Y.Z., V.L. and A.E.M. designed the research. Y.Z. performed the research. Y.Z., V.L. and A.E.M. wrote the manuscript.\\
\noindent \textbf{Competing interests:} The authors declare that they have no competing interests.\\
\noindent \textbf{Data availability:} 
All data needed to evaluate the conclusions in the paper are present in the paper and Supplementary Information. Additional data related to this paper may be requested from the authors.\\
\noindent \textbf{Code availability:} 
The Python and MATLAB code implementing the SBD Algorithm is available at \url{https://github.com/y-z-zhang/SBD}.

\end{document}